\documentclass[a4paper,11pt]{article}
\usepackage{pos}
\usepackage{siunitx}
\usepackage{footnote}
\usepackage{hyperref}
\usepackage{wrapfig}
\usepackage{color-edits}
\addauthor[Pascal]{pascal}{orange}
\usepackage{todonotes}
\usepackage{caption}
\usepackage{subcaption}

\title{The Relevance of Muon Deflections for Neutrino Telescopes}
 \ShortTitle{Muon Deflection}

\author*[a]{Pascal Gutjahr}
\author[a]{Jean-Marco Alameddine}
\author[b]{Alexander Sandrock}
\author[a]{Jan Soedingrekso}
\author[a]{Mirco Hünnefeld}
\author[a]{Wolfgang Rhode}

\affiliation[a]{TU Dortmund University, Department of Physics\\
  Otto-Hahn-Str. 4, 44227 Dortmund, Germany}

\affiliation[b]{University of Wuppertal, Faculty of Mathematics and Natural Sciences,\\
Gaußstraße 20, 42119 Wuppertal, Germany}

\emailAdd{pascal.gutjahr@tu-dortmund.de}
\abstract{
  Large-scale neutrino telescopes have the primary objective to detect and characterize neutrino sources in the universe. 
  These experiments rely on the detection of charged leptons produced in the interaction of neutrinos with nuclei. 
  Angular resolutions are estimated to be better than 1 degree, which is achieved by the reconstruction of muons. 
  This angular resolution is a measure of the accuracy with which the direction of incoming neutrinos can be determined. 
  Since muons can traverse distances of several kilometers through media, the original muon direction can differ from the 
  muon direction inside the detector due to deflections by stochastic interactions and multiple scattering.  

  In this contribution, a recently published study of muon deflections based on the simulation tool PROPOSAL is presented. 
  Muons with various energies are propagated through different media over several distances. Data-Monte-Carlo comparisons as
  well as comparisons to the simulation tools MUSIC and Geant4 are performed. Finally, the impact of muon deflections on 
  large-scale neutrino telescopes is discussed.  
}

\ConferenceLogo{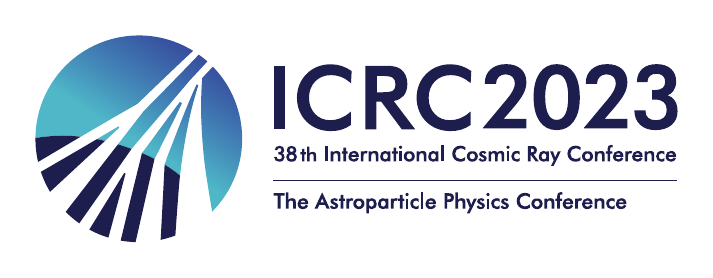}

\FullConference{%
38th International Cosmic Ray Conference (ICRC2023)\\
  26 July - 3 August, 2023\\
  Nagoya, Japan}

\begin{document}
\maketitle

\section{Introduction}
In the universe, neutrino sources are characterized by large-scale neutrino telescopes due to the detection of charged 
leptons produced in interactions of neutrinos with nuclei in the Earth. The utilization of muons enables the best 
opportunity to locate these sources since the large propagation distance of muons even in dense media leads to 
higher angular resolutions in comparison to electrons and taus.
Due to the high distance to these sources, their observations require accurate directional reconstructions 
which are estimated to be better than \SI{1}{\degree} in current neutrino telescopes 
\cite{KM3NeT_Resolution2021, ORCA_Resolution2021, Baikal_Resolution2021, SuperKamiokande_Resolution2008, IceCube_Resolution2021}. 

In these reconstruction algorithms, a muon propagation along a straight line is assumed and deflections of muons are expected to 
be lower than the angular resolutions. However, muons with \si{\peta\electronvolt} energies are able to propagate kilometers 
through media like ice and rock while a multitude of individual interactions occur. With every interaction there is a transfer of energy 
and thus a transfer of momentum, which leads to a small change of direction. All these single deflections accumulate along 
the track and result in a total deflection $\theta_{\mathrm{acc}}$ with respect to the initial muon direction. 
In this proceeding, the impact of this total deflection on the angular resolution of current neutrino telescopes is 
investigated.

\section{Simulation tool PROPOSAL}
The simulations are performed with the open-source Monte-Carlo framework 
PROPOSAL\footnote{\href{https://github.com/tudo-astroparticlephysics/PROPOSAL}{github.com/tudo-astroparticlephysics/PROPOSAL}. The tool can 
be installed with \textit{pip install proposal} or via CMake.} 
\cite{koehne2013proposal,dunsch2018proposal}, which is written in C++ and also available in Python. 
This tool propagates charged leptons and photons through media using state-of-the-art parametrizations and cross-sections and it is 
used in the simulation chain of the IceCube Neutrino Observatory \cite{icecube_proposal}, KM3NeT \cite{km3net_proposal} and CORSIKA 8 \cite{CORSIKA8}.

For muons, the main interaction types bremsstrahlung, photonuclear interaction, electron pair production, ionization and the 
decay are provided. An interaction process is sampled by the given cross-sections. 
The propagation is specified by an initial energy $E_{\mathrm{i}}$, a final energy $E_{\mathrm{f}}$, a propagation distance $d$, a medium 
and energy cuts which are described by an absolute energy cut $e_{\mathrm{cut}}$ and a relative cut $v_{\mathrm{cut}}$. 
These energy cut settings define a minimum energy loss $E_{\mathrm{loss,\,min}}$ by 
\begin{wrapfigure}[12]{r}{5cm}
  \centering 
  \includegraphics[width=4.5cm]{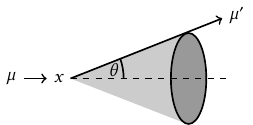}
  \caption{An incoming muon $\mu$ is deflected due to an arbitrary interaction $x$ by the angle $\theta$ resulting to a new 
  direction of the outgoing muon $\mu^\prime$ located on a cone \cite{gutjahr21masters}.}
  \label{fig:3D_angle}
\end{wrapfigure}
\begin{equation}
  E_{\mathrm{loss,\,min}} = \mathrm{min}(e_{\mathrm{cut}}, E\cdot v_{\mathrm{cut}})\,,
\end{equation}
with the energy $E$ before the interaction. An energy loss $E_{\mathrm{loss}} \geq E_{\mathrm{loss,\,min}}$ is treated as a stochastic 
energy loss, $E_{\mathrm{loss}} < E_{\mathrm{loss,\,min}}$ as a continuous energy loss. These cuts are needed since energy losses 
exchanging the massless photon like bremsstrahlung can be arbitrarily small and result in an infinite number of 
interactions without these cuts. Higher cuts lead to less precise propagations but improved runtime performance. 

Deflections along a continuous energy loss are described by multiple scattering. Different multiple scattering parametrizations are available. 
The deflections caused by stochastic interactions 
have recently been implemented in PROPOSAL version $7$ and are available for all relevant muon interaction types 
\cite{Van_Ginneken, GEANT4}. A sketch of a single muon deflection is visualized in Figure~\ref{fig:3D_angle}. 
Further investigations can be found in \cite{gutjahr21masters,gutjahr22deflectionpaper}.

In addition, a scattering multiplier $\zeta$ is introduced for all deflection parametrizations and also multiple scattering. 
This multiplier scales sampled deflection angles $\theta$ by
\begin{equation}
  \theta_{\mathrm{s}} = \zeta \cdot \theta\,. \label{eqn:scat_mult}
\end{equation}
For each interaction type, a separate multiplier can be chosen. On the one hand, this scaling enables further investigations 
of upper deflection limits, on the other hand it can be used to perform an analysis in which a deflection can 
be measured.

A detailed description of PROPOSAL is stated in \cite{koehne2013proposal,dunsch2018proposal, phd_soedingrekso}.
For this proceeding, all simulations are performed with PROPOSAL version $7.6.2$ using the default parametrizations 
and multiple scattering described by Molière \cite{moliere_scattering} and an energy cut of $e_{\mathrm{cut}} = \SI{500}{\mega\electronvolt}$
if not stated otherwise.

\section{Resulting muon deflections}

A single muon deflection depends on the interaction type, the muon energy $E$ and the energy loss $E_{\mathrm{loss}}$. 
For \num{1000} muons propagated from $E_{\mathrm{i}} = \SI{1}{\peta\electronvolt}$ to $E_{\mathrm{f}} = \SI{1}{\tera\electronvolt}$ in ice, 
\SI{95}{\percent} of all deflections occur in the interval of $[\num{2.2e-7}\,\si{\degree},\,\num{1.3e-3}\,\si{\degree}]$ 
with a median deflection of \SI{3.9e-6}{\degree}. The median propagation distance  
with a \SI{95}{\percent} interval is $d = 16.4^{+24.6}_{-7.3}\,\si{\kilo\meter}$ \cite{gutjahr22deflectionpaper}.

\begin{figure}
  \centering
  \includegraphics[width=0.9\textwidth]{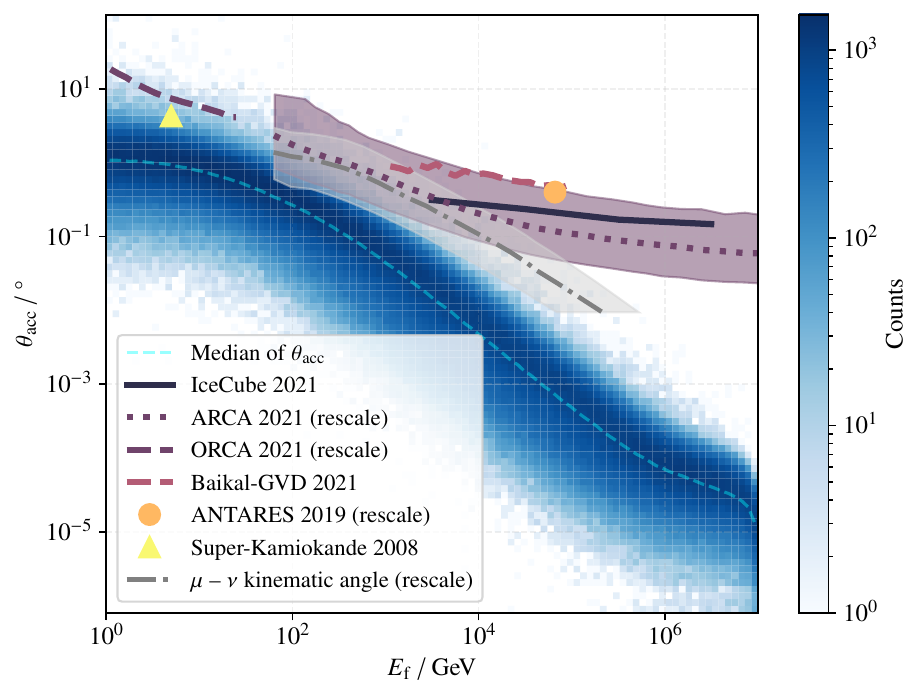}
  \caption{The accumulated muon deflection $\theta_{\mathrm{acc}}$ is shown in dependence of the final muon energy $E_{\mathrm{f}}$ for 
  \num{e6} muons propagated through ice. Angular resolutions of detectors are shown for IceCube~\cite{IceCube_Resolution2021}, ARCA~\cite{KM3NeT_Resolution2021}, 
  ORCA~\cite{ORCA_Resolution2021}, Baikal-GVD~\cite{Baikal_Resolution2021}, ANTARES~\cite{ANTARES_Resolution2019} and Super-Kamiokande~\cite{SuperKamiokande_Resolution2008}. The kinematic production 
  angle between a neutrino and the produced muon is taken from~\cite{KM3NeT_Resolution2021}. Some resolutions are stated as a function of the 
  neutrino energy, to compare these with the muon energy, a rescaling is performed by applying the average 
  energy transfer to the nucleus~\cite{GANDHI199681}. 
  The same simulations are performed in water resulting in a deviation of the median deflection of less than \SI{1}{\percent}.}
  \label{fig:final_plot}
\end{figure}

In the following, the impact of the muon deflection on the angular resolution of muon and neutrino detectors is studied.
For this, the accumulated muon deflection $\theta_{\mathrm{acc}}$ between the initial muon direction 
and the muon direction after a propagation is simulated. 
First, \num{e6} muons are propagated from $E_{\mathrm{i}} = \SI{10}{\peta\electronvolt}$ to $E_{\mathrm{f}} = \SI{1}{\giga\electronvolt}$
in ice. This is presented in Figure~\ref{fig:final_plot}. The lower the final muon energy, the larger the accumulated deflection. 
The profile of the entire distribution has a width of approximately three orders of magnitude. At \SI{1}{\giga\electronvolt}, 
the median deflection is about \SI{1}{\degree}, at very high energies of \SI{1}{\peta\electronvolt} it is about 
\SI{e-4}{\degree}. Furthermore, angular resolutions of neutrino telescopes are added. At energies above \SI{10}{\tera\electronvolt}, 
there is no impact of the muon deflection on the resolution. The resolution of Baikal-GVD~\cite{Baikal_Resolution2021}
which is on the order of \SI{1}{\degree}, is stated for energies larger than \SI{1}{\tera\electronvolt} and not affected, similar 
for ANTARES~\cite{ANTARES_Resolution2019}. However, the reconstruction performance of ARCA~\cite{KM3NeT_Resolution2021} and  
ORCA~\cite{ORCA_Resolution2021} are both impacted for energies below \SI{10}{\tera\electronvolt}. This is also the case for 
Super-Kamiokande~\cite{SuperKamiokande_Resolution2008}. For IceCube~\cite{IceCube_Resolution2021}, the resolution mentioned here might be 
impacted by some outliers between \SI{1}{\tera\electronvolt} and \SI{10}{\tera\electronvolt}. Additionally, the kinematic scattering angle 
between the neutrino and the produced muon is added, which is larger than the deflection~\cite{KM3NeT_Resolution2021}.

\begin{figure}
  \centering
  \begin{subfigure}[t]{0.48\textwidth}
      \centering
      \includegraphics[width=\textwidth]{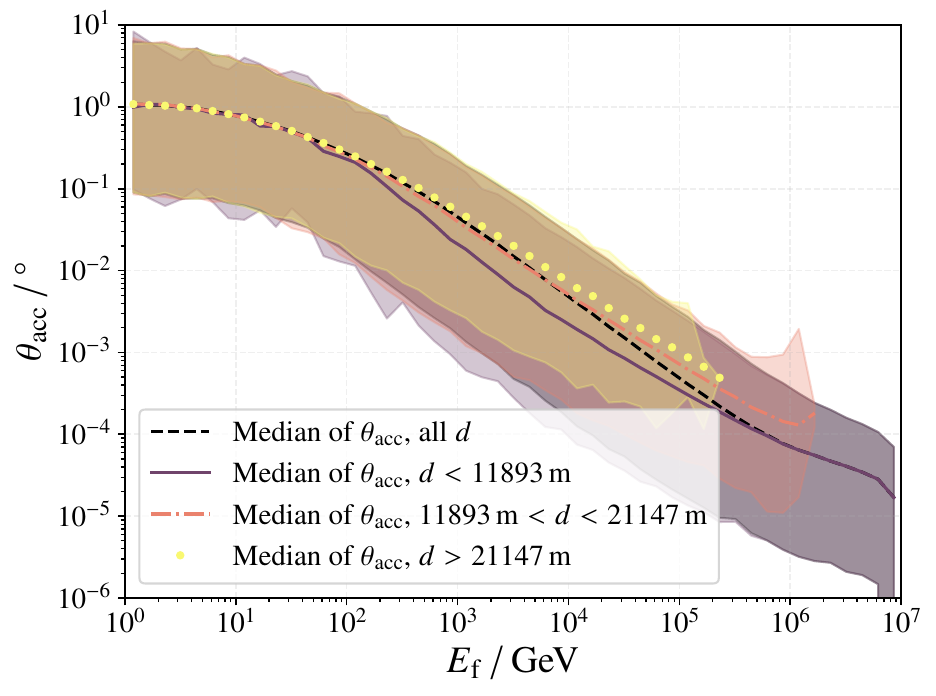}
      \caption{The simulation data set of Figure~\ref{fig:final_plot} is divided into three distance quantiles that 
      contain an equal number of events. Presented are four different medians of the accumulated 
      deflection with \SI{99}{\percent} 
      . For low distances, lower deflections occur between \SI{100}{\giga\electronvolt} and 
      \SI{1}{\peta\electronvolt}. Large distances lead to slightly higher deflections.}
      \label{fig:propagation_distance}
  \end{subfigure}
  \hfill
  \begin{subfigure}[t]{0.48\textwidth}
      \centering
      \includegraphics[width=\textwidth]{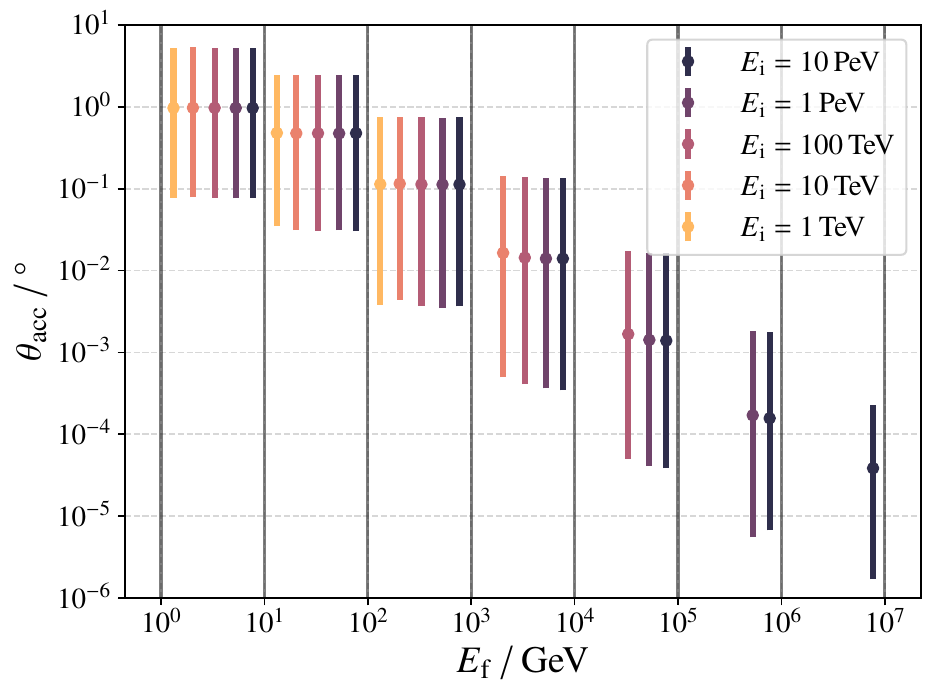}
      \caption{The median deflection $\theta_{\mathrm{acc}}$ with the \SI{99}{\percent} intervals are shown for different initial energies 
      $E_{\mathrm{i}}$ and logarithmic binned final energies $E_{\mathrm{f}}$. Each simulation set contains \num{e6} muons. 
      The grey vertical lines indicate seven intervals of final energies used to determine the medians.
      Medians and \SI{99}{\percent} intervals overlap.}
      \label{fig:defl_e_i} 
  \end{subfigure}
  \caption{The impact of the propagation distance $d$ and the initial muon energy $E_{\mathrm{i}}$ on the deflection 
  are investigated in dependence of the final muon energy $E_{\mathrm{f}}$. The distance impacts the deflection only 
  slightly and the initial muon energy has nearly no impact.}
  \label{fig:prop_and_defl_e_i}
\end{figure}

Furthermore, the deflection is investigated in dependence of the propagation distance $d$ for the same 
simulation data set. All distances are divided into three 
intervals that contain an equal number of events. The medians of accumulated deflections are presented for 
these three intervals and all distances in Figure~\ref{fig:propagation_distance}. At high final energies, large 
distances do not occur since the muon loses energy continuously during the propagation which results in a maximum distance 
that can be reached for a specific setting of an initial and final energy. For distances $d < \SI{11893}{\meter}$, the 
median deflection is shifted to lower 
angles for energies between \SI{100}{\giga\electronvolt} and \SI{1}{\peta\electronvolt}. For distances $d > \SI{21147}{\meter}$, 
angles are slightly higher in comparison to the median of all events. 
All in all, there is only a small impact of the propagation distance on the deflection.

On the experimental side of a neutrino telescope, the initial muon energy is not known. Hence, the deflection 
is analyzed in dependence of different initial energies. In total, \num{e6} muons are propagated through 
ice for five different initial energies $E_{\mathrm{i}} \in 
[\SI{1}{\tera\electronvolt},\,\SI{10}{\tera\electronvolt},\,\SI{100}{\tera\electronvolt},\,\SI{1}{\peta\electronvolt},\,\SI{10}{\peta\electronvolt}]$.
Median deflections and \SI{99}{\percent} intervals for seven final energy bins are shown in Figure~\ref{fig:defl_e_i}. 
All medians per final energy bin and even the intervals are similar. From this follows that the accumulated 
muon deflection is independent of the initial muon energy.

\begin{figure}
  \centering 
  \begin{subfigure}{0.48\textwidth}
    \centering
    \includegraphics[width=\textwidth]{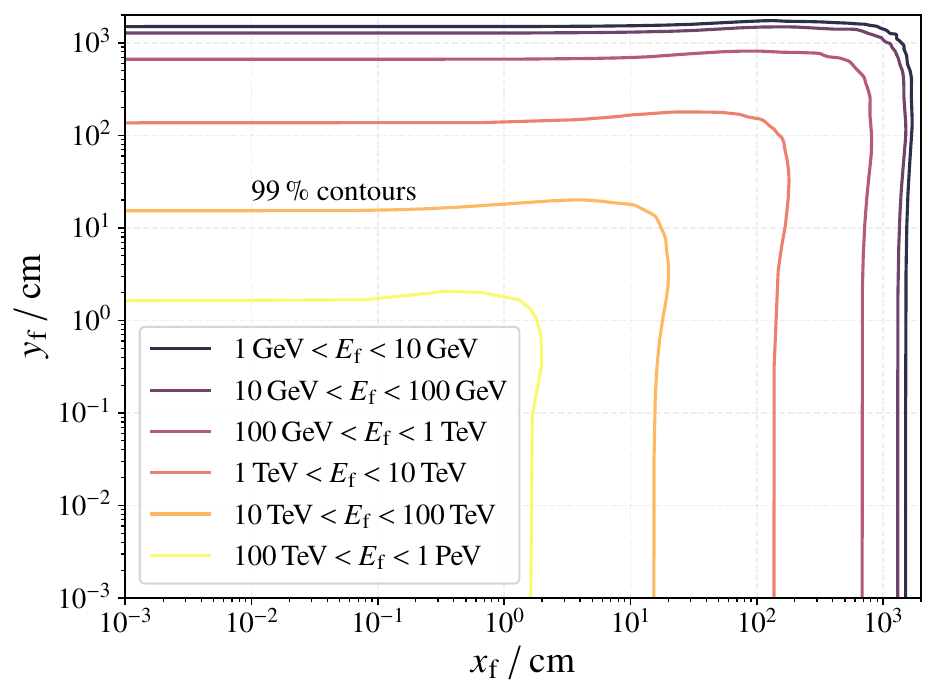}
    \caption{\num{e6} muons with an initial energy of $E_{\mathrm{i}} = \SI{1}{\peta\electronvolt}$ are propagated 
    through ice along the $z$--axis. The \SI{99}{\percent} contours of the lateral distributions are presented for different final 
    energies $E_{\mathrm{f}}$. Since the profiles are circular and isotropic, the absolute values of $x_{\mathrm{f}}$ and  
    $x_{\mathrm{f}}$ are shown. The lower the final energy, the larger the 
    lateral displacement.}
    \label{fig:lateral_profile}
  \end{subfigure}
  \hfill 
  \begin{subfigure}{0.48\textwidth}
    \centering
    \includegraphics[width=\textwidth]{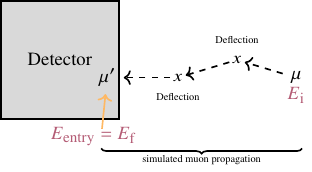}
    \caption{The path of a muon to the detector is sketched. The final muon energy $E_{\mathrm{f}}$ corresponds to the 
    entry energy $E_{\mathrm{entry}}$, that can be reconstructed by a detector. Since the muon deflection is nearly 
    independent of the initial muon energy and the distance, $E_{\mathrm{entry}}$ can be used to estimate the muon deflection before the 
    detector entry.}
    \label{fig:defl_unc_sketch}
  \end{subfigure}
  \caption{The lateral muon profile and a sketch to estimate the muon deflection as a systematic uncertainty for angular reconstructions of 
  muon and neutrino detectors are shown.}
  \label{fig:defl_e_i_and_unc_sketch}
\end{figure}

Additionally, the lateral displacement of the muons is studied in Figure~\ref{fig:lateral_profile}. The 
$E_{\mathrm{i}} = \SI{1}{\peta\electronvolt}$ simulation data set is used and the final energies are divided into six 
energy bins. The \SI{99}{\percent} contours represent a circular and isotropic distribution of the muon 
displacements. The lower the final energy, the larger the displacement. 
This is caused by the fact that larger angles also cause wider displacements. The displacements in the smallest energy 
interval of $\SI{1}{\giga\electronvolt} < E_{\mathrm{f}} < \SI{10}{\giga\electronvolt}$ are smaller than $\SI{1}{\meter}$.

Finally, the accumulated muon deflection depends primarily on the final muon energy. The influence of the propagation distance and the initial muon energy 
are negligible. This allows using the reconstructed muon energy inside a detector to estimate the 
muon deflection before the detector entry. This can serve as an estimation of the systematic uncertainty due to muon deflections for angular reconstruction methods. 
This principle is sketched in Figure~\ref{fig:defl_unc_sketch}. A parametrization of the median deflection as a function 
of the final muon energy is given in \cite{gutjahr22deflectionpaper}.

\section{Summary}
Using the simulation tool PROPOSAL, the accumulated muon deflection and its impact on the resolution of angular reconstruction methods 
in muon and neutrino detectors are investigated. One conclusion is that the lower the final muon energy, the larger is the 
the resulting deflection. Resolutions of Super-Kamiokande, ARCA and ORCA are affected by outliers of the deflection distributions. 
Further studies show that the deflection is nearly independent of the initial muon energy and the propagation distance. 
From this follows that the reconstructed muon energy inside a detector can be used to estimate the deflection of the 
muon before the detector entry. This can serve as an estimation of a systematic uncertainty in reconstruction algorithms.

In addition, resolutions shown in this proceeding are from 2021 and older. Due to new machine learning techniques, the 
reconstruction methods improve and lead to more precise angular resolutions. Thus, the impact of the deflection can 
increase in the future.

\section{Discussion: Data-Monte-Carlo Deviations}
\begin{figure}
  \centering 
  \begin{subfigure}{0.48\textwidth}
    \centering 
    \includegraphics[width=\textwidth]{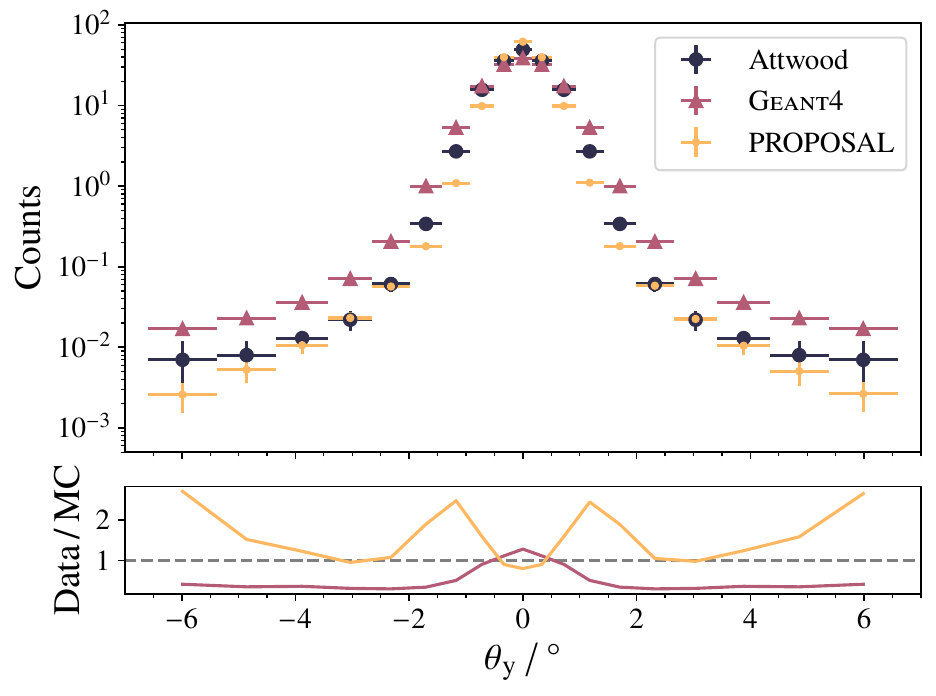}
    \caption{Muons with $E_{\mathrm{i}} = \SI{199}{\mega\electronvolt}$ propagating through \SI{109}{\milli\meter} of liquid $\mathrm{H}_2$ are 
    shown. Deflections are underestimated by PROPOSAL. Data are taken from \cite{attwood_2006}.}
    \label{fig:attwood}
  \end{subfigure}
  \hfill 
  \begin{subfigure}{0.48\textwidth}
    \centering 
    \includegraphics[width=\textwidth]{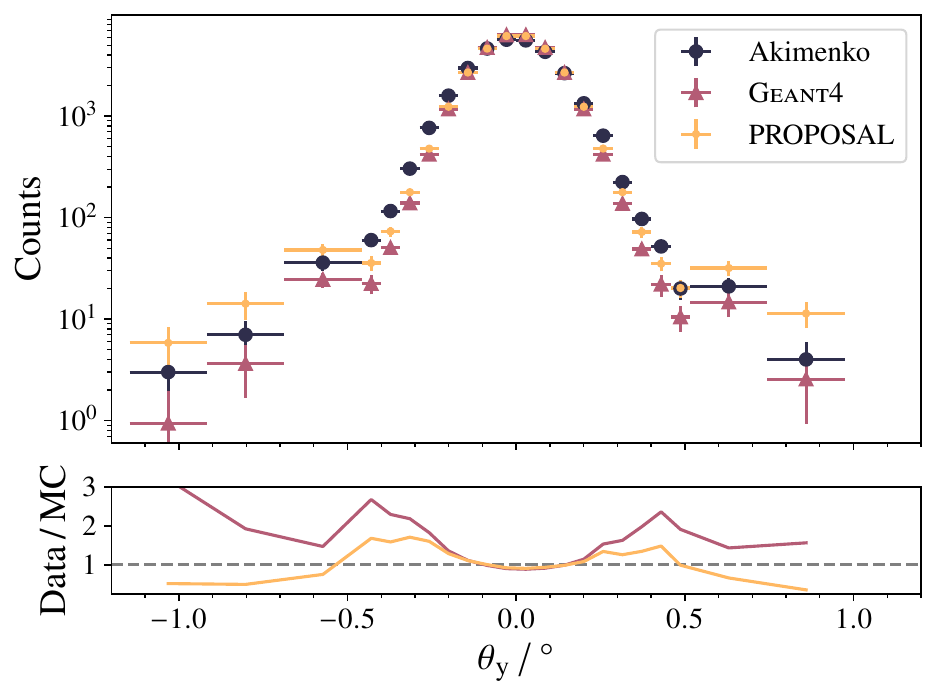}
    \caption{Muons with $E_{\mathrm{i}} = \SI{7.3}{\giga\electronvolt}$ propagating through \SI{144}{\milli\meter} of copper are 
    shown. Larger deflections are overestimated by PROPOSAL. Data are taken from \cite{akimenko_1984}.}
    \label{fig:akimenko}
  \end{subfigure}
  \caption{Two comparisons between measured data, PROPOSAL simulations and the simulation tool \textsc{Geant4} \cite{GEANT4} 
  are presented. In PROPOSAL, \num{100} simulations for each comparison are performed. The means with the standard deviations are shown. 
  A relative energy cut $v_{\mathrm{cut}} = \num{e-5}$ is used. 
  In both comparisons, deviations between the simulated deflection and the measured data occur. Comparisons are taken 
  from \cite{gutjahr22deflectionpaper}.}
  \label{fig:data_MC}
\end{figure}

In recent studies, mismatches between measured data and Monte-Carlo simulations are observed for two different setups \cite{gutjahr22deflectionpaper}.
First, the deflection of muons with $E_{\mathrm{i}} = \SI{199}{\mega\electronvolt}$ passing \SI{109}{\milli\meter} of 
liquid $\mathrm{H}_2$ is measured 
and compared to the results of PROPOSAL \cite{attwood_2006}. Second, muons with 
$E_{\mathrm{i}} = \SI{7.3}{\giga\electronvolt}$ passing \SI{1.44}{\centi\meter} of copper are investigated \cite{akimenko_1984}.
For $E_{\mathrm{i}} = \SI{199}{\mega\electronvolt}$, larger angles are underestimated. For $E_{\mathrm{i}} = \SI{7.3}{\giga\electronvolt}$, 
larger angles are overestimated. Deviations up to a factor of three are observed. So far, these are the only measurements of muon deflections. 
For both measurements, simulations have also been performed with the simulation tool \textsc{Geant4} \cite{GEANT4}, which demonstrates 
similar deviations. 
The comparisons are displayed in Figure~\ref{fig:data_MC}.

By now, muon deflections have been measured only for low energies and short distances with respect to neutrino telescopes. Even at these, 
mismatches occur. Thus, a validation of the simulated muon deflection requires further measurements. These need to be performed at higher energies of 
\si{\tera\electronvolt} to \si{\peta\electronvolt} energies and large scale distances on the order of kilometers. The scattering multiplier 
introduced in Eq.~\ref{eqn:scat_mult} can be used to perform a statistical analysis to measure such deflections. In addition, 
simulations with applied multipliers can serve to estimate upper limits of the muon deflection to testify, which multiplier would 
affect a specific angular reconstruction method.

All in all, the resulting muon deflections simulated in this proceeding extend over several orders of magnitude. Hence, 
possible deviations by a factor of $2-3$ are negligible in first order and the drawn conclusions should remain 
unaffected.

\acknowledgments
This work has been supported by the Deutsche Forschungsgesellschaft (DFG) and the Lamarr institute. Pascal Gutjahr acknowledges the financial 
support by the German Academic Exchange Service (DAAD).

\bibliographystyle{JHEP}
\bibliography{lit.bib}
%% Full authors list (ONLY FOR COLLABORATIONS)
%\clearpage
%\section*{Full Authors List: \Coll\ Collaboration}
%
%\noindent \textbf{Note comment afterwards:} Collaborations have the possibility to provide an authors list in xml format which will be used while generating the DOI entries making the full authors list searchable in databases like Inspire HEP. \\
%
%\scriptsize
%\noindent
%first.author$^1$, 
%second.author$^2$, 
%third.author$^3$ % .... more names
%and 
%last.author$^{n}$ \\
%
%\noindent
%$^1$first.affiliation.
%$^2$second.affiliation. % .... more affiliation
%$^{m}$last.affiliation.

\end{document}